# All-optical 3-input OR and 2-input AND/NIMPLY logic gates in a linear planar three-core optical fiber coupler

J. P. T. Rodrigues[1], F. L. B. Martins[2], and J. C. do Nascimento[3,*]
*[1, 2, 3] Department of Teleinformatic Engineering, Universidade Federal do Ceará, Fortaleza, Brazil.*

---

**Abstract**: Most all-optical logic processing devices reported in the literature rely on nonlinear effects to achieve logical discrimination, which typically results in increased implementation complexity, higher optical power requirements, and limited scalability over larger optical circuits. Here, we present the numerical modeling and analysis of two all-optical logic gates based on a fully linear, dispersion-free planar symmetric three-core optical fiber coupler. The proposed devices operate with low-powered amplitude-modulated pulses and require no nonlinear materials, material doping, or optoelectronic components. By exploiting only the fiber geometry and coupled-mode theory, we demonstrate two logic processing functionalities: a 3-input OR gate and a configurable gate that performs either the AND or the NIMPLY operation based on a control signal. Logical discrimination is measured using a contrast-based performance metric, showing clear separation between logic levels for all input combinations. The simplicity of the proposed linear structure, combined with its scalability and compatibility with different fiber technologies, favor compact implementation and integration into larger photonic signal processing and optical computing systems.

---

## 1. Introduction

The all-optical processing of ultrafast pulses has attracted increasing interest as a means to overcome the speed and power limitations imposed by optical-electro-optical conversion in communication and signal processing systems [1 – 3]. In this context, optical fiber-based devices are particularly attractive for their low losses, compactness, and compatibility with existing photonic infrastructures. Among these, multicore fiber architectures with multiple coupled cores have shown great potential for applications in all-optical switching, multiplexing, and logic processing [4 – 8].

Most all-optical logic devices reported in the literature rely on nonlinear effects or additional optoelectronic components to achieve logical discrimination [9 – 20]. While this approach may provide increased switching functionality, it often also increases device complexity, power consumption, and sensitivity to fabrication imperfections, which limits scalability and integration into larger circuits. All-optical devices operating in a purely linear regime remain comparatively rare, despite their potential advantages in terms of simplicity, robustness, and ease of integration [4 – 6].

An alternative strategy is avoiding nonlinear mechanisms altogether by using only linear coupling effects in carefully designed multicore fiber structures. In this approach, logical processing can be accomplished through the geometry of the fiber and the controlled exchange of optical power between cores, allowing compact devices to perform nontrivial logic operations with low computational and experimental complexity.

In this work, we numerically model two linear all-optical-fiber-based logic devices using a planar symmetric three-core optical fiber coupler operating in a linear dispersion-free pulse propagation regime. One implements a 3-input OR logic gate, the other implements a configurable two-input logic gate that performs either the AND or the NIMPLY operation depending on the logical value of a control signal. These results further demonstrate that meaningful all-optical logic processing can be achieved with purely linear multicore fiber designs.

## 2. Planar Symmetric Three-Core Optical Fiber Coupler (PSTC Coupler)

A PSTC coupler consists of three identical cores arranged in a planar, symmetric, and equidistant configuration, as shown in Figure 1. The cores are labeled as 1, 2, and 3, and the optical signals propagating through them are denoted as **A**, **B**, and **C**, respectively. The geometric parameters Λ and d represent the core-to-core separation and the core diameter, respectively. The coupling coefficient between adjacent cores is denoted by κ. The separation between nonadjacent cores, 2Λ, is sufficiently large to prevent any optical interaction between them. Like this, signal **A** (in core 1) interacts and trades power only with signal **B**; signal **B** (in core 2) interacts and trades power with both signals **A** and **C**; and signal **C** (in core 3) interacts and trades power only with signal **B**. Signals **A** and **C** do not interact with each other (therefore trading no power).

**Figure 1:** Schematic representation of a PSTC coupler with length z, core diameter d and core separation Λ. (a) 3D model and (b) cross-sectional schematic.

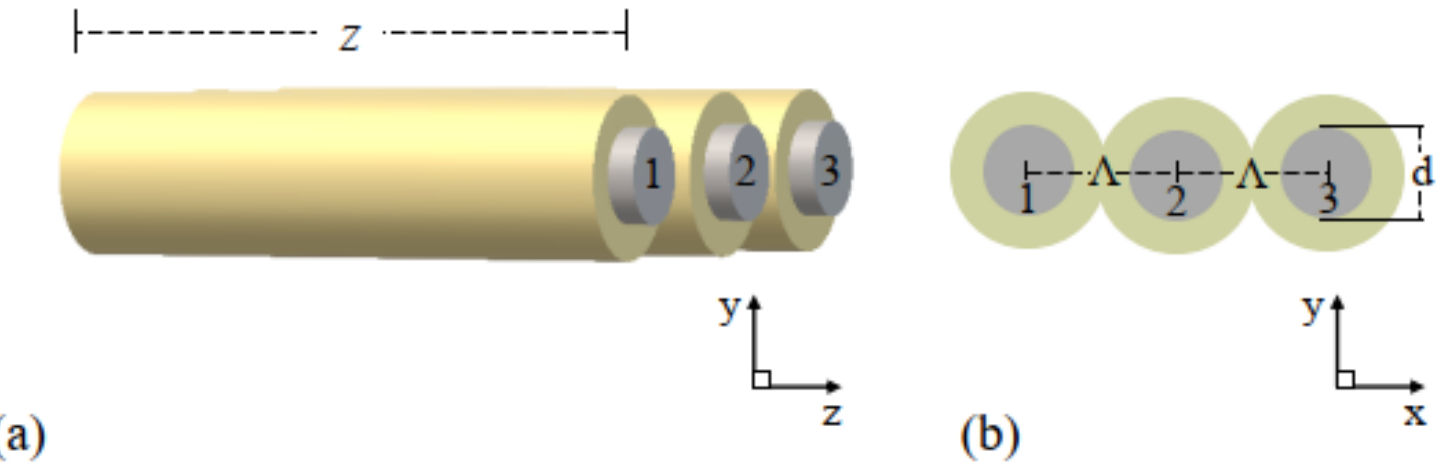

Our device is designed to operate in a linear and dispersion-free pulse propagation regime. By employing low-powered optical pulses and a coupler length much shorter than both the dispersion and nonlinear lengths, nonlinear effects and dispersive broadening can be safely neglected [4 – 6, 21, 22]. Under these conditions, the temporal profile of the pulses remains unchanged during propagation, and only the optical field amplitudes evolve along the coupler's length, z. As a result, the dynamics of the system can be accurately described by a set of coupled-mode equations that account solely for linear evanescent coupling between adjacent cores:

$$\frac{\partial a_1}{\partial z} = i\kappa a_2, \tag{1}$$

$$\frac{\partial a_2}{\partial z} = i\kappa a_1 + i\kappa a_3, \tag{2}$$

$$\frac{\partial a_3}{\partial z} = i\kappa a_2, \tag{3}$$

where $a_1(z,t)$, $a_2(z,t)$ and $a_3(z,t)$ denote the complex optical field amplitudes associated with signals **A**, **B**, and **C**, respectively. The $t$ parameter will be omitted for simplification (it becomes irrelevant in a linear dispersion-free setup since there are no time-dependent changes in the pulse).

Arranging the field amplitudes into the vector $\vec{a}(z) = [a_1(z), a_2(z), a_3(z)]^T$, Equations (1), (2) and (3) can be written in matrix form as:

$$\frac{d\vec{a}(z)}{dz} = A\vec{a}(z), \tag{4}$$

where the coupling matrix is given by:

$$A = \begin{bmatrix} 0 & i\kappa & 0 \\ i\kappa & 0 & i\kappa \\ 0 & i\kappa & 0 \end{bmatrix}. \tag{5}$$

Matrix A is diagonalizable and admits the spectral decomposition $A = PDP^{-1}$, where:

$$P = \begin{bmatrix} 1 & 1 & -1 \\ -\sqrt{2} & \sqrt{2} & 0 \\ 1 & 1 & 1 \end{bmatrix}, \tag{6}$$

$$D = \begin{bmatrix} -\sqrt{2}\kappa i & 0 & 0 \\ 0 & \sqrt{2}\kappa i & 0 \\ 0 & 0 & 0 \end{bmatrix}, \quad \text{and} \tag{7}$$

$$P^{-1} = \frac{1}{4}\begin{bmatrix} 1 & -\sqrt{2} & 1 \\ 1 & \sqrt{2} & 1 \\ -2 & 0 & 2 \end{bmatrix}. \tag{8}$$

The solution of Equation 4 can then be expressed using the matrix exponential as

$$\vec{a}(z) = Pe^{zD}P^{-1}\vec{a}(0), \qquad (9)$$

where $Pe^{zD}P^{-1}$ is the optical field transfer matrix, and $\vec{a}(0)$ and $\vec{a}(z)$ are the input and output field amplitude vectors, respectively. $e^{zD}$ is given by:

$$e^{zD} = \begin{bmatrix} e^{-\sqrt{2}z\kappa i} & 0 & 0 \\ 0 & e^{\sqrt{2}z\kappa i} & 0 \\ 0 & 0 & 1 \end{bmatrix}. \qquad (10)$$

Since $\kappa$ and $z$ always appear as a product, we can simplify the analysis by reparameterizing them as $\theta = z\kappa$. Like this, the optical field transfer matrix, $Pe^{zD}P^{-1}$, can be expressed as:

$$Pe^{zD}P^{-1}(\theta) = \begin{bmatrix} \frac{\cos(\sqrt{2}\theta)+1}{2} & \frac{i\sqrt{2}\sin(\sqrt{2}\theta)}{2} & \frac{\cos(\sqrt{2}\theta)-1}{2} \\ \frac{i\sqrt{2}\sin(\sqrt{2}\theta)}{2} & \cos(\sqrt{2}\theta) & \frac{i\sqrt{2}\sin(\sqrt{2}\theta)}{2} \\ \frac{\cos(\sqrt{2}\theta)-1}{2} & \frac{i\sqrt{2}\sin(\sqrt{2}\theta)}{2} & \frac{\cos(\sqrt{2}\theta)+1}{2} \end{bmatrix}. \qquad (11)$$

This transfer matrix enables the calculation of the output pulse amplitudes in all cores for any input condition in the PSTC coupler, as described in Equation 9. The $\theta = z\kappa$ reparameterization simplifies these calculations and allows multiple physical realizations of the same logical functionality, since multiple valid combinations of $z$ and $\kappa$ can lead to the same value of $\theta$.

## 3. Pulse Modulation and Logic Evaluation

Our device operates in a linear dispersion-free regime, so the temporal shape of the pulses is preserved during propagation and only their amplitudes and phases vary along the fiber. Since phase variation is identical in all three cores (due to simultaneous input, identical propagation speeds, and equal fiber lengths), optical signal processing can only be achieved through the exchange of optical field amplitudes between coupled cores. So, we modulate the signals solely in amplitude, directly mapping logical information onto the pulse amplitude.

We consider hyperbolic secant-shaped pulses with normalized reference amplitude $a_{ref} = 1$ and a modulation parameter $\varepsilon$ that varies within the range of $0 < \varepsilon < 1$, generalizing both pulse amplitude modulation (PAM) and on-off keying (OOK) schemes, with OOK emerging as a limiting case for $\varepsilon \to 1$. In this setup, a bit 0 input pulse is given by $a_0(0,t) = (1-\varepsilon)\text{sech}\,(t)$, whereas a bit 1 input pulse is given by $a_1(0,t) = (1+\varepsilon)\text{sech}\,(t)$. The reference pulse is $a_{ref}(0,t) = \text{sech}\,(t)$.

In demodulation, logical discrimination is performed by comparing the output pulse amplitude $a_{out}$ with the normalized reference level. Pulses with $a_{out} > 1$ are interpreted as bit 1, whereas pulses with $a_{out} < 1$ are interpreted as bit 0. To quantify how distinguishable the output pulses are from the reference, we use the contrast ratio metric, defined as

$$C_R = 20\log_{10}(a_{out}). \quad (12)$$

A positive contrast ($C_R > 0$) corresponds to a logical bit 1, whereas a negative contrast ($C_R < 0$) corresponds to a logical bit 0.

In practical implementations, noise may originate from several stages of the optical link: the optical source, the fiber medium, photodetectors, and even possible amplification stages preceding the device. Rather than modeling several specific noise mechanisms, we adopt $|C_R| > 0.3$ dB (which corresponds to an eye opening of approximately 0.6 dB) as the minimum threshold required to distinguish the output logic signals from noise. This value should be interpreted as a conservatively low, experimentally measurable limit at which the two logical levels first become distinguishable by high-quality photoreceivers, rather than as a practical operating margin [5, 6]. Higher contrast ratios or enhanced detection strategies (e.g., coherent receivers or nonlinear thresholding) would be required in realistic implementations. Still, this approach allows the performance analysis to remain independent of any particular experimental configuration while ensuring a consistent logic discrimination criterion.

## 4. Results

Given Equations 9 and 11, our experiment consisted of exploring the $(\varepsilon, \theta)$ parametric space (within the ranges of $0 < \theta < \pi/\sqrt{2}$ and $0 < \varepsilon < 1$, with step sizes of 0.0025 and 0.01, respectively) in a truth table while identifying its corresponding logic functionalities. Only configurations in which $|C_R| > 0.3$ dB for all input combinations were retained, ensuring reliable logical discrimination according to the criterion defined in Section 3.

We identified several high-performance logic functions. The two most interesting results are described in Table 1, which summarizes the logic functions implemented at each output core. Here, outputs $\mathbf{Y}_1$, $\mathbf{Y}_2$, and $\mathbf{Y}_3$ correspond to cores 1, 2, and 3, respectively.

**Table 1:** Logical functions implemented in each output core.

| Core – Output | Device 1 | Device 2 | |
|---|---|---|---|
| | | $\mathbf{C} = 0$ | $\mathbf{C} = 1$ |
| Core 1 – $\mathbf{Y}_1$ | – | AND | – |
| Core 2 – $\mathbf{Y}_2$ | 3-input OR | – | – |
| Core 3 – $\mathbf{Y}_3$ | – | – | NIMPLY |

### 4.1. First device: 3-input OR logic gate

The first device implements a 3-input OR logic gate at the output of core 2 ($\mathbf{Y}_2$). This functionality is observed for two different modulation schemes: pulse amplitude modulation (PAM) with $\varepsilon_1 = 0.3$ and $\theta_1 = 0.585$ rad, and on–off keying (OOK) as a limiting case with $\varepsilon_{1'} = 1$ and $\theta_{1'} = 0.675$ rad.

To illustrate pulse propagation, we simulated the insertion of $100$ fs pulses into the device for each combination of logic values in the truth table. Figures 2 and 3 show the temporal amplitudes of the output pulses for both modulation schemes. For all input combinations, the output of core 2 exhibits amplitudes well above the reference level whenever at least one input is equal to '1', while remaining below the reference level only when all inputs are '0' (which confirms the correct implementation of the 3-input OR function ($\mathbf{A} + \mathbf{B} + \mathbf{C}$)).

Table 2 shows the truth tables for the 3-input OR logic gate implemented using both modulation schemes. For each input combination (logical values of signals $\mathbf{A}$, $\mathbf{B}$, and $\mathbf{C}$), the table lists logical outputs (logical values of signals $\mathbf{Y}_1$, $\mathbf{Y}_2$, and $\mathbf{Y}_3$), the optical output amplitudes ($\mathrm{a_{out}}$), and the corresponding contrast ratios ($\mathrm{C_R}$). Under PAM modulation, the minimum contrast ratio observed is $|\mathrm{C_R}| = 1.1570$ dB. Under OOK, the minimum value is $|\mathrm{C_R}| = 1.2449$ dB. Both values are significantly above the adopted threshold of $0.3$ dB, indicating robust logical discrimination with a comfortable noise margin.

**Figure 2:** Temporal amplitudes of the output pulses of the PAM modulated 3-input OR logic gate.

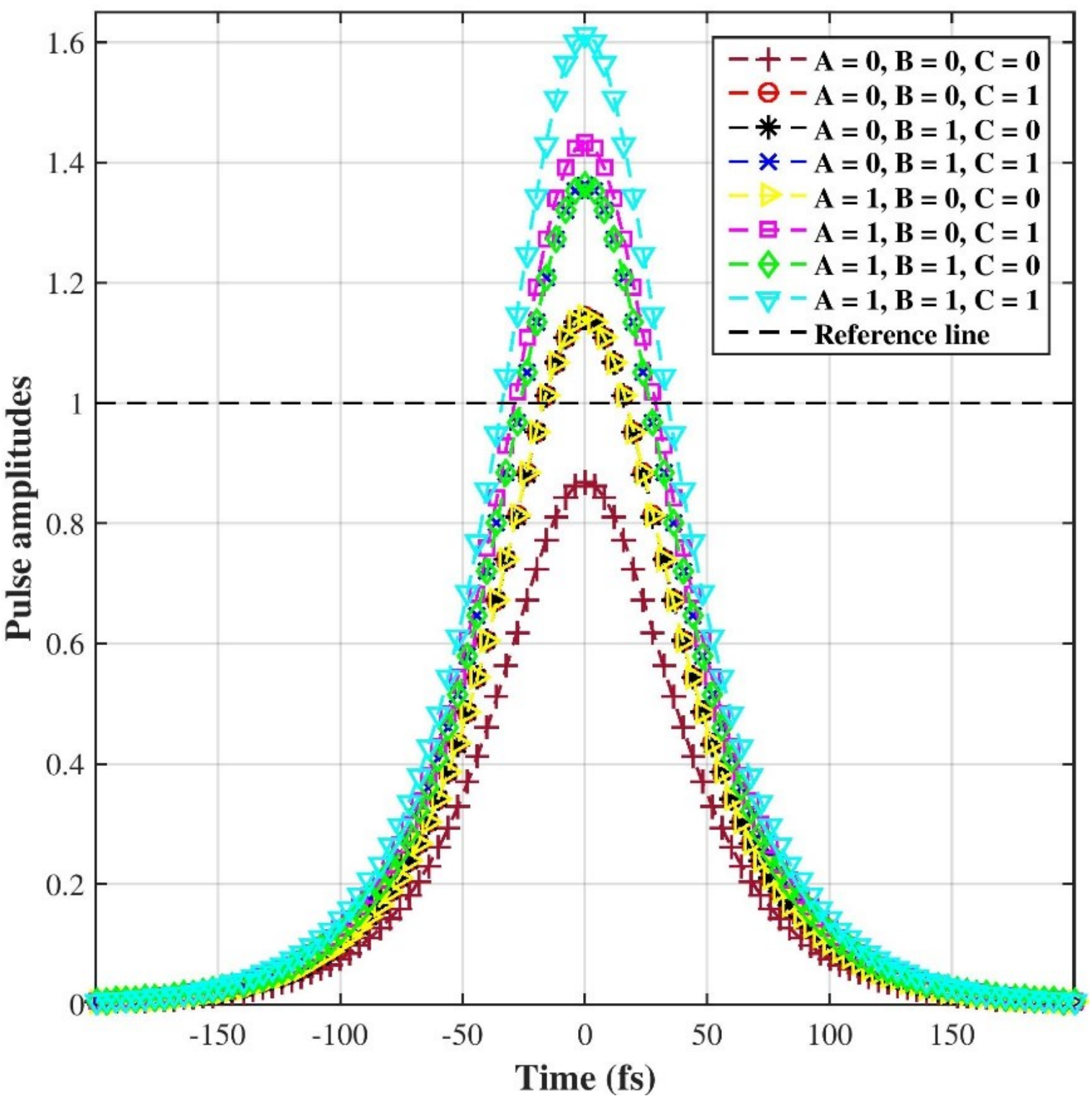

**Figure 3:** Temporal amplitudes of the output pulses of the OOK modulated 3-input OR logic gate.

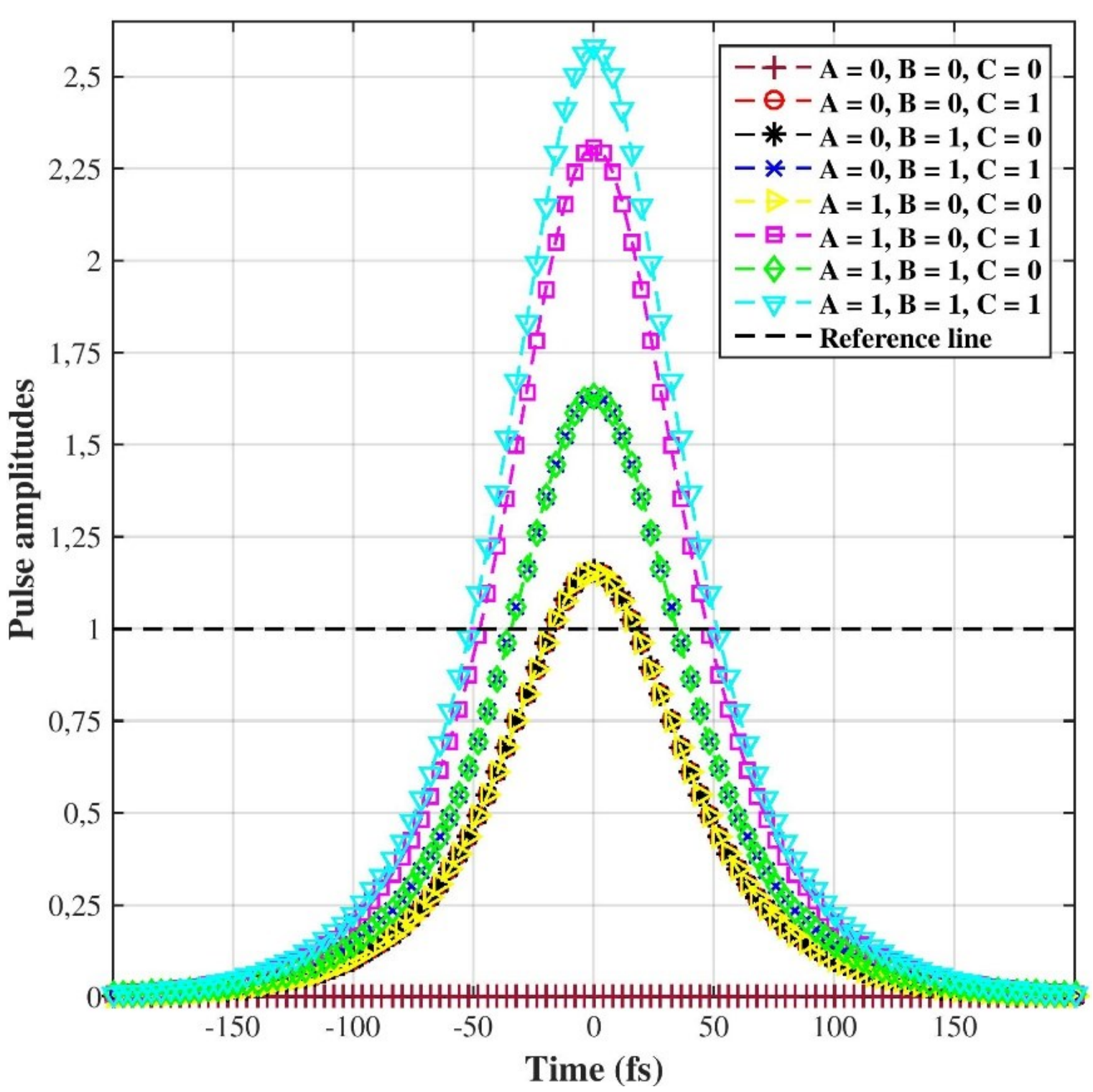

**Table 2:** Truth table, amplitudes and contrast ratio values of the output pulses for the 3-input OR gate.

| Logic Gate | Logic Inputs | | | Outputs (Core 2) | | | | |
|---|---|---|---|---|---|---|---|---|
| | | | | Logic Output | PAM Modulation | | OOK Modulation | |
| | **A** | **B** | **C** | $\mathbf{Y_2}$ | $a_{out}$ | $C_R$ (dB) | $a_{out}$ | $C_R$ (dB) |
| | 0 | 0 | 0 | 0 | 0.8662 | –1.2176 | 0 | – |
| | 0 | 0 | 1 | 1 | 1.1438 | +1.1668 | 1.1541 | +1.2449 |
| | 0 | 1 | 0 | 1 | 1.1425 | +1.1570 | 1.1559 | +1.2583 |
| | 0 | 1 | 1 | 1 | 1.3631 | +2.6904 | 1.6334 | +4.2619 |
| | 1 | 0 | 0 | 1 | 1.1438 | +1.1668 | 1.1541 | +1.2449 |
| | 1 | 0 | 1 | 1 | 1.4339 | +3.1302 | 2.3082 | +7.2655 |
| | 1 | 1 | 0 | 1 | 1.3631 | +2.6904 | 1.6334 | +4.2619 |
| | 1 | 1 | 1 | 1 | 1.6142 | +4.1593 | 2.5815 | +8.2373 |

### 4.1.1. Tolerance analysis

To gauge robustness, we analyzed the tolerance ranges for variations in $\varepsilon_1$ and $\theta_1$ for which the first device maintains acceptable performance. Figure 4 shows the regions in the $(\varepsilon_1, \theta_1)$ parametric space for which the $|C_R| \geq 0.3$ dB condition is satisfied. For PAM, centered at $\varepsilon_1 = 0.3$ and $\theta_1 = 0.585$ rad, $\theta_1$ can vary within $0.475 < \theta_1 < 0.85$ (corresponding to a tolerance of 18.80%), and $\varepsilon_1$ can vary within $0.2333 < \varepsilon_1 < 1$ (a tolerance of 23.33%). For OOK, the ranges are $0.5825 < \theta_{1'} < 0.725$ (7.41%) and $0.65 < \varepsilon_{1'} < 1$ (35%). The first device exhibits substantial tolerance to both fabrication-related variations (implicit in $\theta = z\kappa$) and modulation-related deviations, which reinforces the practical viability of the proposed linear design.

**Figure 4:** Region of occurrence region and contrast ratios highlighting multiple intervals of 0.3 dB in the $(\varepsilon, \theta)$ parametric space.

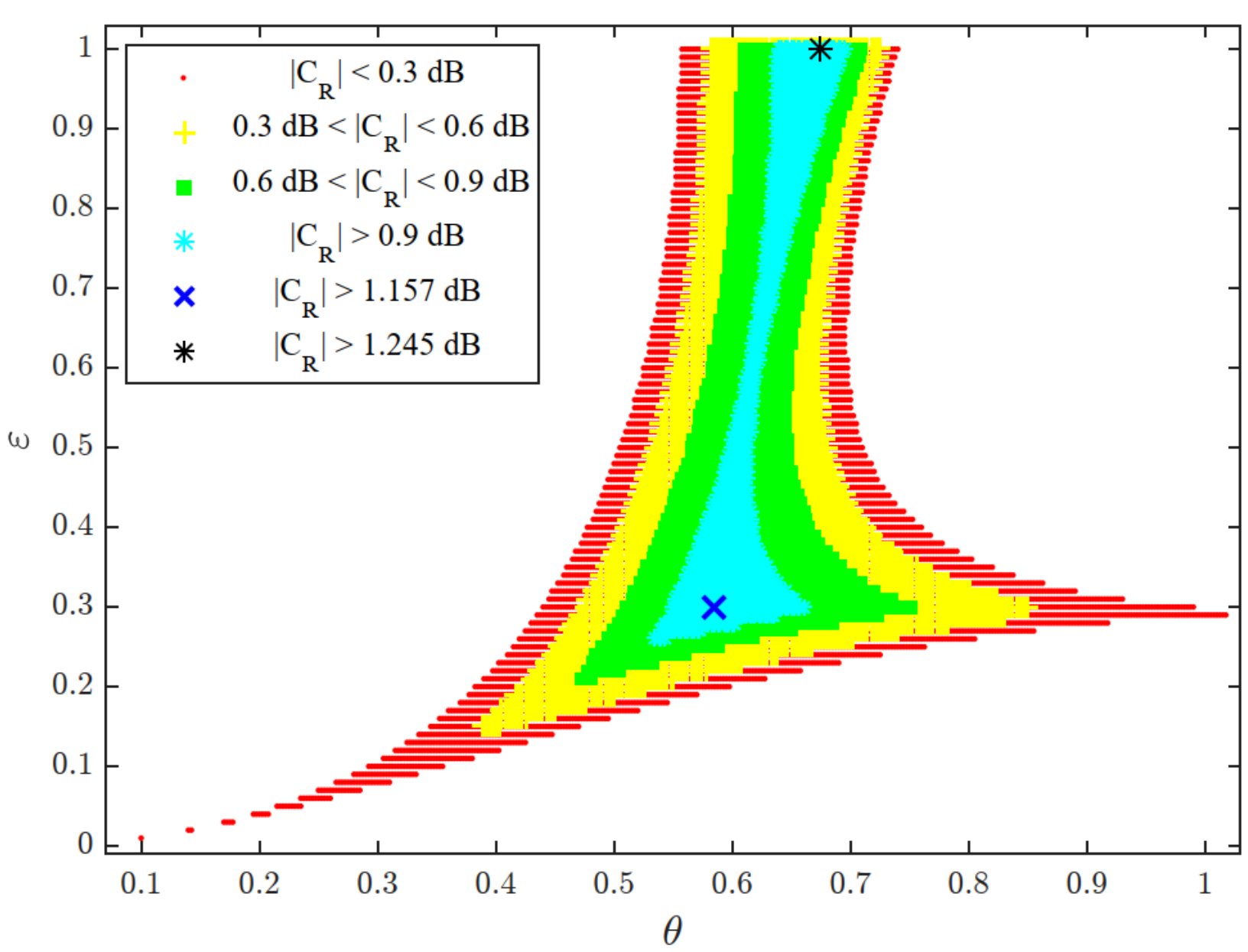

### 4.2. Second device: configurable AND/NIMPLY logic gate

The second device implements a configurable AND/NIMPLY logic gate, with the AND and NIMPLY operations being realized at different output cores based on a control signal. This functionality is observed for $\varepsilon_2 = 0.31$ and $\theta_2 = 0.945$ rad, where input signal $\mathbf{C}$ (in core 3) acts as a control signal while $\mathbf{A}$ and $\mathbf{B}$ (in cores 1 and 2, respectively) serve as the logical operands. When the control signal is set to $\mathbf{C} = 0$, the device performs the 2-input AND operation at the output of core 1 ($\mathbf{Y}_1 = \mathbf{AB}$). For $\mathbf{C} = 1$, the device performs the NIMPLY operation at the output of core 3 ($\mathbf{Y}_3 = \overline{\mathbf{A}}\mathbf{B}$, also represented as $\mathbf{A} \nrightarrow \mathbf{B}$ or $\overline{\mathbf{A} \rightarrow \mathbf{B}}$).

Figures 5 and 6 show the temporal amplitudes of the output pulses for the AND and NIMPLY operations, respectively. In both cases, the output amplitudes clearly exceed or remain below the reference level according to the corresponding truth tables (which confirms the correct implementation of both logical operations).

Tables 3 and 4 show the truth tables, the output amplitudes, and the corresponding contrast ratios for the both gates. For the AND gate, the minimum contrast ratio observed is $|C_R| = 0.4368$ dB. For the NIMPLY gate, the minimum value is $|C_R| = 0.4367$ dB. While closer to the adopted threshold, these still remain safely above the minimum requirement of 0.3 dB, ensuring reliable logical discrimination.

**Figure 5:** Temporal amplitudes of the output pulses of the 2-input AND gate.

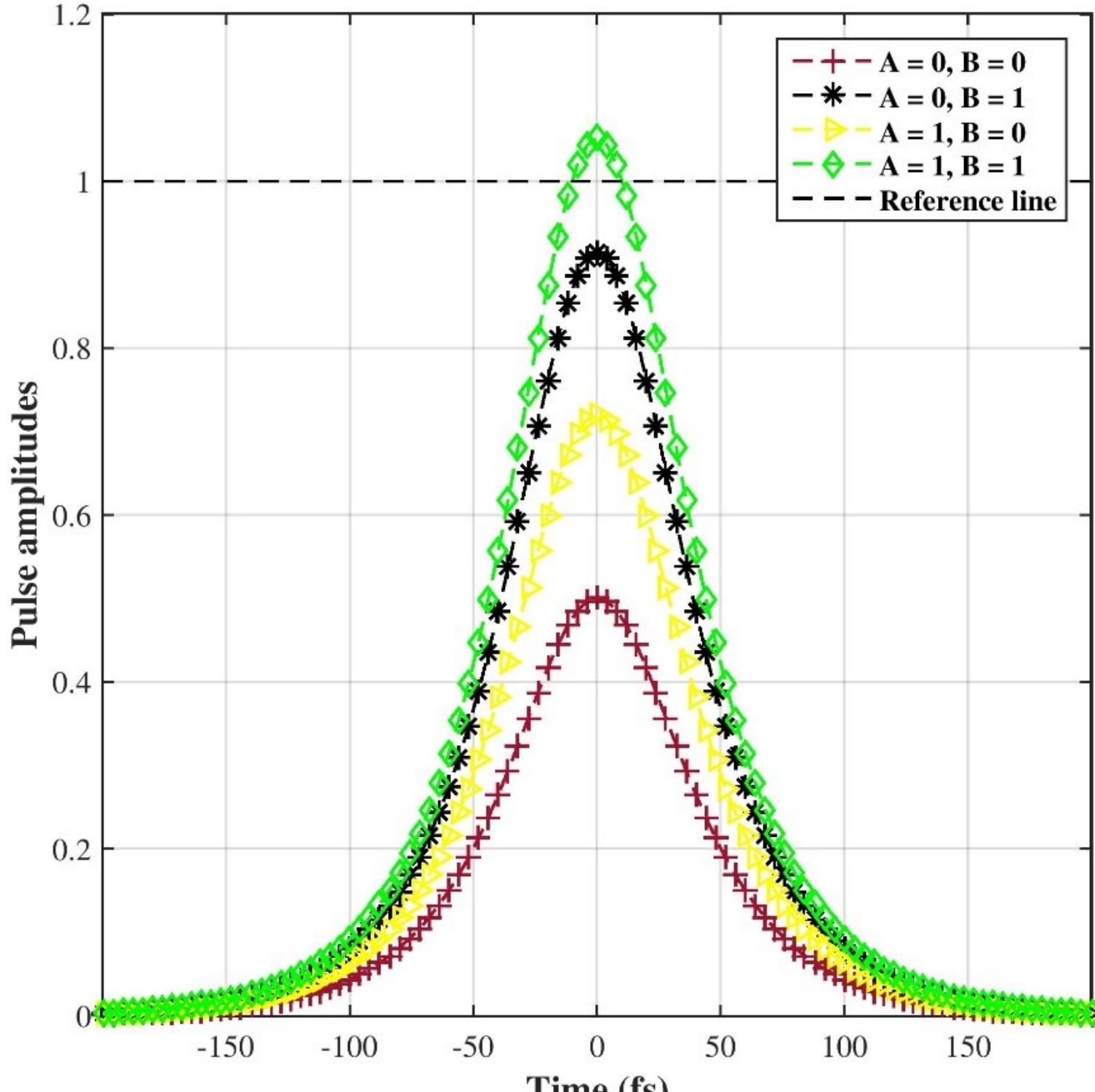

**Table 3:** Truth table, amplitudes and contrast ratio values of the output pulses for the 2-input AND gate.

| Logic Gate | Inputs | | | Outputs (Core 1) | | |
|---|---|---|---|---|---|---|
| | **A** | **B** | **C** | $\mathbf{Y_1}$ | $a_{out}$ | $C_R$ (dB) |
| AND | 0 | 0 | 0 | 0 | 0.5009 | –6.0052 |
| | 0 | 0 | 1 | 0 | 0.9151 | –0.7704 |
| | 0 | 1 | 0 | 0 | 0.7206 | –2.8485 |
| | 0 | 1 | 1 | 1 | 1.0516 | +0.4368 |

**Figure 6:** Temporal amplitudes of the output pulses of the NIMPLY gate.

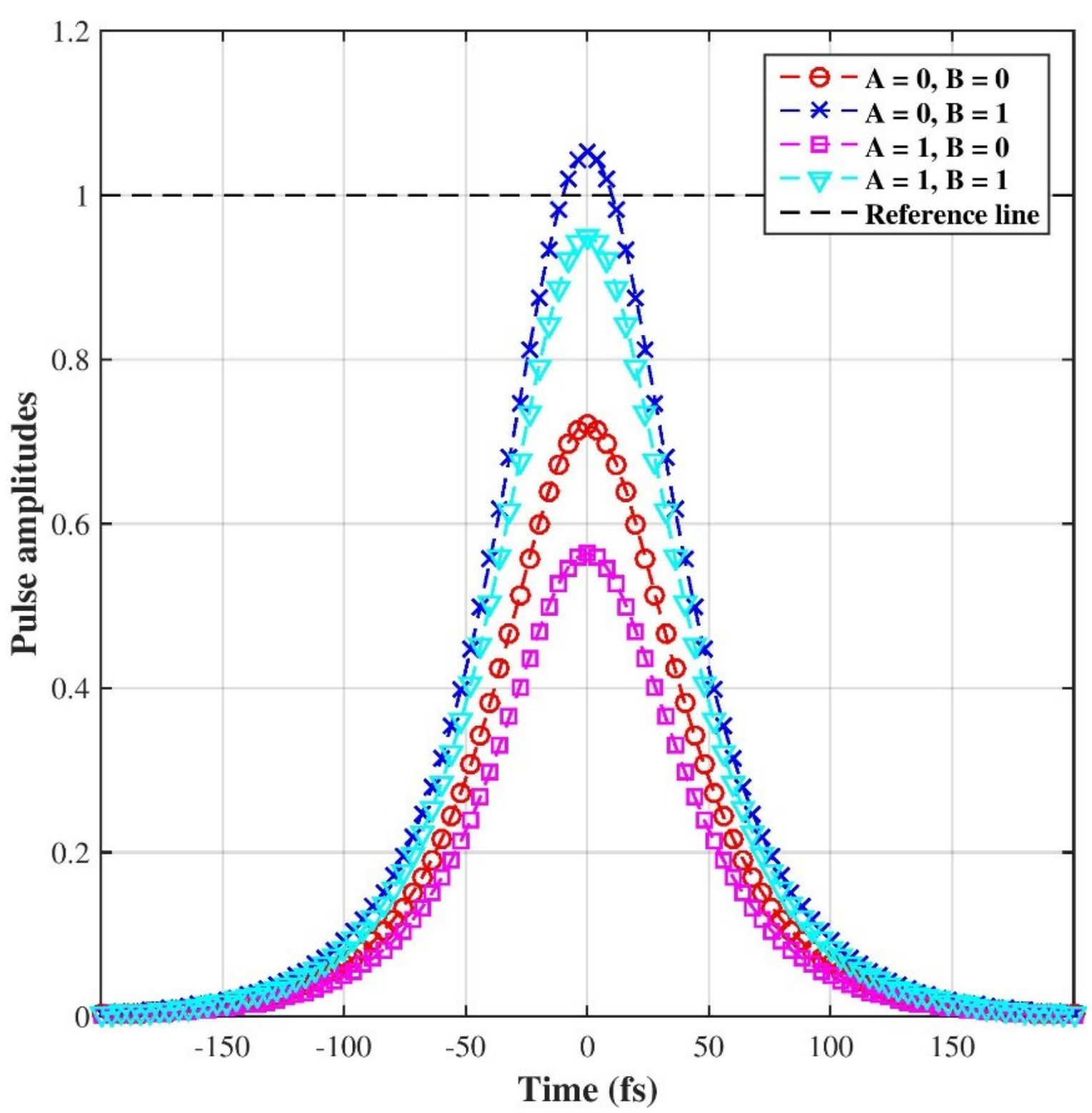

**Table 4:** Truth table, amplitudes and contrast ratio values of the output pulses for the 2-input NIMPLY gate.

| Logic Gate | Inputs | | | Outputs (Core 3) | | |
|---|---|---|---|---|---|---|
| | **A** | **B** | **C** | $\mathbf{Y_3}$ | $a_{out}$ | $C_R$ (dB) |
| | 1 | 0 | 0 | 0 | 0.7206 | –2.8465 |
| | 1 | 0 | 1 | 1 | 1.0516 | +0.4367 |
| | 1 | 1 | 0 | 0 | 0.5637 | –4.9790 |
| | 1 | 1 | 1 | 0 | 0.9510 | –0.4368 |

### 4.2.1. Tolerance analysis

As in section 4.1.1, we analyzed the tolerance ranges for variations in $\varepsilon_2$ and $\theta_2$ for which the second device maintains acceptable performance. Figure 7 shows the regions in the $(\varepsilon_2, \theta_2)$ parametric space for which the $|C_R| \geq 0.3$ dB condition is satisfied for both the AND and NIMPLY operations. For a fixed value of $\varepsilon_2 = 0.31$, $\theta_2$ can vary within the range $0.9 < \theta_2 < 0.98$ (corresponding to a tolerance of 3.70%). Similarly, for a fixed value of $\theta_2 = 0.945$ rad, $\varepsilon_2$ can vary within the range $0.30 < \varepsilon_2 < 0.33$ (a tolerance of 3.23%).

While this margin is theoretically sufficient for logic discrimination (with the minimum contrast ratio of $|C_R| = 0.4368$ dB corresponding to an eye opening of approximately 0.8736 dB), it may be substantially reduced in practice by small fabrication-induced deviations, detector noise,

and other experimental imperfections. Practical implementations may require high-sensitivity photodetectors with high SNR, low noise, and high-resolution ADCs, or alternative detection strategies such as coherent optical receivers and nonlinear thresholding techniques [5, 6].

**Figure 7:** Region of occurrence and contrast ratios highlighting the 0.3 dB threshold in the $(\varepsilon, \theta)$ parametric space.

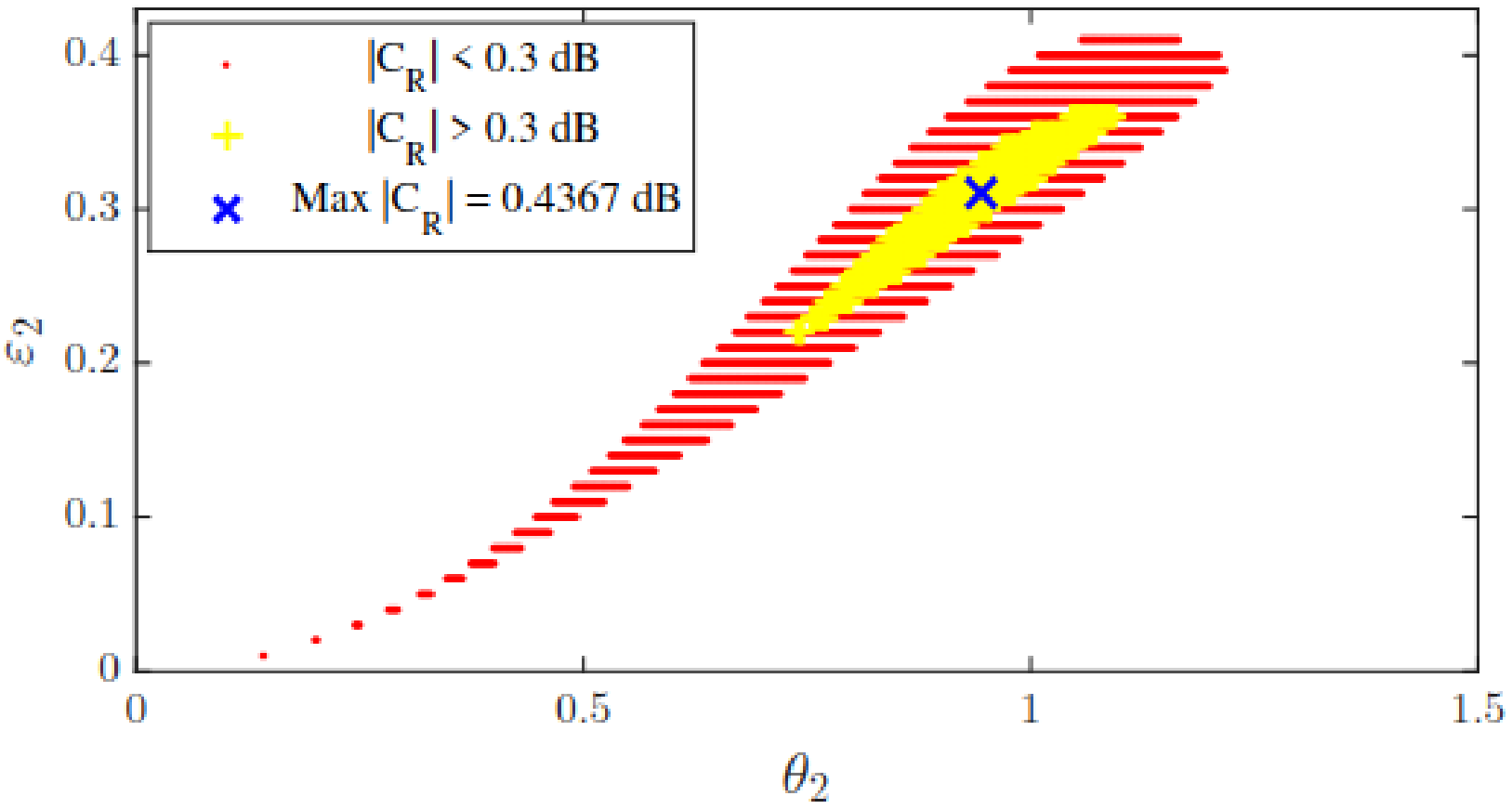

## 5. Discussion

Our results demonstrate that complex all-optical logic functionalities can be achieved using purely linear optical fiber-based devices. By encoding logical information directly into the optical field amplitude and exploiting controlled linear coupling between adjacent cores, the proposed devices implement both a combinational logic gate (the 3-input OR gate, which is the combination/cascading of two basic 2-input OR gates) and a configurable conditional-logic gate (the AND/NIMPLY gate) without relying on nonlinear effects or optoelectronic components.

Unlike most approaches to all-optical digital logic processing devices, which depend on nonlinear (or even highly nonlinear) materials and effects to achieve logic processing, the present linear dispersion-free architecture achieves reliable operation through appropriate coupler design and modulation parameter selection. This challenges the widespread assumption that nonlinear effects are a necessary requirement for optical logic processing in fiber-based platforms. While linear dispersion-free systems do not provide the intrinsic gain, bistability, nor the increased switching capabilities (and variability of results) often seen in nonlinear systems, they more than make up for it with their simplicity of modeling, reduced physical, energetical and computational costs, and the significantly reduced dispersion, distortion, and phase shifts suffered during pulse propagation (which increases the potential for successful device concatenation [5, 6]).

### 5.1. Experimental Implementation

From a manufacturing perspective, the planar geometry offers practical advantages over alternative three-core configurations. PSTC couplers are inherently simpler to fabricate and integrate than triangular symmetric three-core structures, as they involve a reduced number of geometric constraints and a two-dimensional core arrangement. This simplification relaxes fabrication tolerances, lowers production complexity, and facilitates integration with other photonic components. A more detailed discussion of these aspects can be found in reference by Martins et al. [5], where the comparative advantages of planar implementations are analyzed in the context of linear all-optical logic devices.

The experimental realization of the proposed devices is governed by a set of physical and operational constraints, with the most critical being the precise control of the modulation and coupling parameters: $\varepsilon$ and $\theta$. Optical pulses must be injected synchronously into all three cores of the coupler. In practice, this requires the three input pulses to enter the device simultaneously (without relative phase shifts) and with accurately defined amplitudes of $(1-\varepsilon)$ for bit 0 and $(1+\varepsilon)$ for bit 1. Asynchronous input conditions inhibit the coupling-mediated redistribution of optical power, thereby suppressing the emergence of the intended logical functionalities, while amplitude errors may degrade performance or even induce unwanted logic-state transitions, particularly in the configurable AND/NIMPLY gate. Likewise, accurate tuning of $\theta$ demands strict control of both the propagation length z (during fabrication, path equalization, and fiber cleaving) and the coupling coefficient $\kappa$ (which is determined by the fiber technology, geometry, and operating pulse wavelength).

To ensure that the system operates within the linear dispersion-free regime assumed here, both the propagation length $z$ and the input pulse power $P_0$ must be carefully constrained: $z$ must be much shorter than both the dispersion length $L_D$ and the nonlinear length $L_{NL}$; and $P_0$ must be sufficiently low to prevent the onset of nonlinear effects. Under these conditions, pulse broadening and waveform distortion are negligible, and the coupling coefficient $\kappa$ can be accurately treated as constant over the signal bandwidth. Residual spectral dependence of $\kappa$ may still introduce minor amplitude variations for broadband pulses, but these effects can be mitigated through appropriate wavelength selection and pulse shaping [23].

### 5.2. Applicability

An important aspect of the proposed approach is its generality with respect to the underlying physical platform. The analysis does not rely on a specific material, wavelength, or fiber technology. Instead, it depends only on the fulfillment of the coupling and propagation conditions discussed above. Once a PSTC coupler is fabricated and its geometric parameters are defined, $\kappa$ can be determined either experimentally or via established empirical formulas [24, 25]. The required device length for a given logical functionality then follows directly from the relation $z = \theta/\kappa$. As a result, different physical implementations sharing the same $\theta$ are expected to exhibit the same logical behavior, even if their absolute dimensions, guiding mechanisms and structural designs markedly differ.

As an example, we'll analyze the fabrication requirements using two fundamentally different technological platforms: conventional silica fiber (CSF) and photonic crystal fiber (PCF). For the CSF PSTC coupler, we'll adopt: $\kappa = 0.3312$ rad/m, $d = 9$ μm, $\Lambda = 11$ μm, $\lambda = 1.55$ μm, $\beta_2 = 0.02$ ps$^2$/m, and $\gamma = 0.002 W^{-1}m^{-1}$, as in [5]. For the PCF PSTC coupler, we'll adopt: $\kappa = 87.266$ rad/m, $d = 2$ μm, $\Lambda = 4.44$ μm, $\lambda = 1.55$ μm, $\beta_2 = -0.047$ ps$^2$/m, and $\gamma = 0.0032 W^{-1}m^{-1}$, as in [13].

Using the CSF platform, the PAM 3-input OR gate ($\varepsilon_1 = 0.3$, $\theta_1 = 0.585$ rad) requires a coupler length of $z_1 = \theta_1/\kappa = 1.766$ m. At this length, linear dispersion-free pulse propagation is satisfied for $L_D = L_{NL} = 10z = 17.66$ m, peak power $P_0$ below $(\gamma L_{NL})^{-1} = 28.31$ W, and pulse width $T_0$ above $\sqrt{L_D|\beta_2|} = 0.59$ ps, corresponding to an ideal non-overlapping bit rate of $B_R = T_0^{-1} = 1.68$ Tbit/s. Under the same conditions, the AND/NIMPLY gate ($\varepsilon_2 = 0.31$, $\theta_2 = 0.945$ rad) requires $z_2 = \theta_2/\kappa = 2.853$ m, with $L_D = L_{NL} = 28.53$ m, $P_0$ below 17.53 W, $T_0$ above 0.75 ps, yielding $B_R = 1.33$ Tbit/s. These values should be interpreted as theoretical feasibility bounds rather than optimized engineering targets.

Using the PCF platform, the PAM 3-input OR gate requires only $z_1 = 6.704$ mm. In this case, linear dispersion-free operation is ensured for $L_D = L_{NL} = 67.04$ mm, $P_0$ below 4.66 kW, and $T_0$ above 56.13 fs, yielding an ideal bit rate of $B_R = 17.81$ Tbit/s. Likewise, the AND/NIMPLY gate requires $z_2 = 1.083$ cm, with $L_D = L_{NL} = 10.83$ cm, $P_0$ below 2.885 kW, $T_0$ above 71.35 fs, yielding $B_R = 14.02$ Tbit/s.

These results demonstrate the broad adaptability of the proposed approach, which can be tailored to distinct technological scenarios according to priorities such as compactness, manufacturability, cost, integration level, or achievable data rate. Platforms with stronger inter-core coupling, such as PCFs, enable highly compact millimeter-scale devices with ultrahigh potential bit rates, making them attractive for dense photonic integration and short-reach ultrafast processing. CSFs, despite requiring meter-scale device lengths, offer mature fabrication processes, lower costs, mechanical robustness, and straightforward compatibility with existing telecom infrastructure.

### 5.3. Comparison with Previous Works

Most prior works addressing all-optical logic devices rely on nonlinear effects or interference-based mechanisms to achieve digital logic processing [9 – 11, 13, 26]. In contrast, our approach operates entirely in a linear and dispersion-free regime, where logic arises solely from geometry-controlled coupling and amplitude encoding, without requiring nonlinearities, phase control, or active optoelectronic elements. Relative to prior linear multicore fiber implementations [4], it offers reduced structural complexity and device length, simpler fabrication, and expanded logical functionality. To the best of our knowledge, this is the first demonstration of a NIMPLY logic gate implemented in a fully linear optical-fiber-based device. This functionality is particularly relevant because NIMPLY operations are directly associated with reversible and conditional logic architectures, while remaining relatively uncommon even in nonlinear optical implementations.

A particularly relevant point of comparison concerns the 3-input OR logic gate. In a previous work based on planar three-core PCF couplers operating in the nonlinear regime, the 3-input OR function was obtained through soliton propagation and Kerr-induced nonlinear effects [13]. In that approach, logic processing relied on self- and cross-phase modulation, dispersion engineering, and the numerical solution of the nonlinear Schrödinger equation, requiring precise control of pulse energy, temporal width, and phase. By contrast, the 3-input OR gate demonstrated here operates in a linear dispersion-free regime, with the logic processing emerging solely from the controlled linear coupling between adjacent cores and amplitude modulation of low-powered pulses. This represents a significant conceptual simplification, as it removes the need for soliton formation, nonlinear phase accumulation, and phase-sensitive control. This is also, to the best of our knowledge, the first demonstration of such functionality operating entirely in a linear regime while supporting both PAM and OOK modulation formats. Compared to [13], the present device not only eliminates nonlinear effects entirely, but also achieves significantly improved performance, including substantially larger contrast ratios and shorter device lengths under equivalent platform conditions.

From a practical perspective, the linear implementation considerably reduces experimental complexity and power requirements, and eliminates numerical and physical issues commonly associated with nonlinear propagation, such as pulse breakup and sensitivity to parameter fluctuations. These results demonstrate that even logic functions traditionally associated with nonlinear fiber optics, such as multi-input operations, can be realized within a fully linear multicore fiber architecture. These comparisons indicate that the proposed linear three-core coupler architecture provides a competitive and scalable alternative for all-optical logic processing, particularly in applications where simplicity, robustness, and integration into larger linear photonic systems are prioritized over extreme switching speeds or optical gain.

## 6. Conclusion

In this work, we present the numerical design and analysis of two all-optical logic devices based on a planar symmetric three-core optical fiber coupler operating in a linear dispersion-free pulse propagation regime. By encoding logical information directly in the optical field amplitude and exploiting controlled linear coupling between adjacent cores, the proposed devices implement a 3-input OR logic gate and a configurable 2-input AND/NIMPLY gate, both exhibiting clear logical discrimination according to a contrast-based metric.

A key result of this study is the demonstration that multi-input and configurable all-optical logic functionalities (traditionally associated with nonlinear fiber optics) can be achieved using purely linear architectures. In particular, the realization of a 3-input OR gate in the linear regime represents a significant conceptual simplification compared to previous nonlinear implementations, reducing experimental complexity, power requirements, and sensitivity to parameter fluctuations while preserving clear logical discrimination.

The analysis of the required propagation lengths for different fiber technologies further highlights the flexibility of the proposed approach. Depending on the coupling coefficient, the same normalized device operation can be implemented in structures ranging from micrometer-scale photonic crystal fibers to meter-scale conventional silica fibers. This scalability, combined with the simplicity and robustness of linear operation, makes the proposed devices promising building blocks for larger-scale linear optical signal processing and logic architectures.

**Funding Support**

This article was financed in part by the CAPES - Brasil - Finance Code001, and BPI FUNCAP.

**Conflicts of Interest**

The authors declare that they have no conflicts of interest to this work.